\documentclass[12pt,aps,showpacs]{revtex4}
\usepackage{amssymb,amsfonts,epsfig}

\newcommand{\beeq}{\begin{equation}}
\newcommand{\eneq}{\end{equation}}
\newcommand{\be}{\begin{eqnarray}}
\newcommand{\ee}{\end{eqnarray}}
\newcommand{\bpic}{\begin{picture}}
\newcommand{\epic}{\end{picture}}
\newcommand{\bs}{\begin{scriptsize}}
\newcommand{\es}{\end{scriptsize}}

\def\dd{\partial}
\def\la{\raise.16ex\hbox{$\langle$} \, }
\def\ra{\, \raise.16ex\hbox{$\rangle$} }

\def\a{\alpha}

\def\d{\delta}
\def\e{\epsilon}

\def\G{\Gamma}
\def\r{\rho}

\def\D{\Delta}

\def\l{\lambda}

\def\Box{\kern1pt\vbox{\hrule height 1.2pt\hbox{\vrule width 1.2pt\hskip 3pt
   \vbox{\vskip 6pt}\hskip 3pt\vrule width 0.6pt}\hrule height 0.6pt}\kern1pt}
\def\gtwid{\mathrel{\raise.3ex\hbox{$>$\kern-.75em\lower1ex\hbox{$\sim$}}}}
\def\ltwid{\mathrel{\raise.3ex\hbox{$<$\kern-.75em\lower1ex\hbox{$\sim$}}}}

\begin{document}

\vskip -0.2in

\rightline{UFIFT-QG-06-19, CRETE-06-20}

\vskip 0.5in

\title{Quantum Stability of a $w<-1$ Phase of Cosmic Acceleration}

\author{E. O. Kahya}\email{emre@phys.ufl.edu}

\author{V. K. Onemli}\email{onemli@physics.uoc.gr}

\affiliation{$^{\ast}$Department of Physics, University of
Florida, Gainesville, Florida 32611, USA}

\affiliation{$^{\dagger}$Department of Physics, University of
Crete, Heraklion, GR-76003, Greece}

\begin{abstract}
We consider a massless, minimally coupled scalar with a quartic
self-interaction which is released in Bunch-Davies vacuum in the
locally de Sitter background of an inflating universe. It was
shown, in this system, that quantum effects can induce a temporary
phase of super-acceleration, causing a violation of the Weak
Energy Condition on cosmological scales. In this paper, we
investigate the system's stability by studying the behavior of
linearized perturbations in the quantum-corrected effective field
equation at one- and two-loop order. We show that the amplitude of
the quantum-corrected mode function is reduced in time, starting
from its initial classical (Bunch-Davies) value. This implies that
the linear perturbations do not grow, hence the model is stable.
The decrease in the amplitude is in agreement with the system
developing a positive (growing) mass-squared due to quantum
processes. The induced mass, however, remains perturbatively small
and does not go tachyonic. This ensures the stability.
\end{abstract}


\maketitle \vskip 0.2in \vspace{.4cm}

\section{Introduction}

Present cosmological observations \cite{SN} do not exclude the
possibility of an evolving dark energy equation of state $w\equiv
p/\rho$ whose current value is less than minus one
\cite{Caldwell}, i.e., a phase of super-acceleration. Although the
data are consistent with $w=-1$, which can be explained by a
simple cosmological constant, the possibility of $w<-1$ has been
an area of great interest in recent years \cite{research}.

Super-acceleration is difficult to explain with {\it classical}
models on account of the problem with stability \cite{CHT}. One
can achieve models exhibiting~$w<-1$, by postulating scalar
fields, for example. Such models, however, decay irrespective of
how this is achieved. The observed persistence of the universe,
therefore, can only be consistent with a relatively brief
self-limiting phase of super-acceleration. One way to get such a
self-limiting phase, without violating classical stability, is via
quantum effects \cite{PR,Staro,SY,OW1,OW2,BOW,W}. The energy-time
uncertainty principle requires virtual particles to emerge from
the vacuum and then disappear back into it. The inflationary
expansion of spacetime, however, causes the virtual particles
persist longer than the flat spacetime \cite{Ws}. In fact, any
sufficiently long wavelength virtual particle-antiparticle pairs,
which are {\it massless} on the Hubble scale, are pulled apart by
the Hubble flow before they find time to annihilate each other.
Hence, they become real and may persist forever; recalling the
analogy with the Hawking radiation. The rate at which the virtual
particles emerge from the vacuum, on the other hand, is suppressed
by the inverse of the scale factor for conformally invariant
particles. Thus, quantum effects are enhanced during inflation for
particles that are effectively massless (with respect to the
Hubble parameter $H$) and classically conformally noninvariant.
Gravitons and massless minimally coupled (MMC) scalars are unique
in possessing zero mass without having classical conformal
invariance. One is lead, naturally, to a self-limiting quantum
effect in a classically stable theory, such as the MMC scalar with
a quartic self-interaction in the locally de Sitter background of
an inflating universe. The Lagrangian density that describes this
system is\be {\cal L} = -\frac{1}{2}\sqrt{-g}g^{\mu\nu}
\partial_{\mu} \varphi \partial_{\nu} \varphi -
\frac{\lambda}{4!}\sqrt{-g}\varphi^4+{\rm counterterms}\; .
\label{Lag}\ee The dynamical variable in the model is the scalar
field $\varphi(x)$. The metric $g_{\mu\nu}$ is a nondynamical
background which is taken to be a $D$-dimensional locally
de~Sitter geometry. The invariant element can be expressed
conveniently either in comoving or conformal coordinates\be ds^2 =
-dt^2 + e^{2 H t} d\vec{x} \cdot d\vec{x} = a^2(\eta) \Bigl[
-d\eta^2 + d\vec{x} \cdot d\vec{x} \Bigr]\; , \label{metric}\ee
respectively. The conformal factor and the transformation which
relate the two coordinate systems are \be a(\eta) = -{1 \over H
\eta} = e^{H t} \; . \ee  The Hubble constant $H$ is related to
the cosmological constant $\Lambda = (D-1) H^2$. It is the
cosmological constant that drives inflation in the model. The
scalar is a spectator to $\Lambda$-driven (de Sitter) inflation.
We adopt the following notations: $x^\mu = (x^0,\vec x)$,
$x^0\equiv \eta$, $\partial_\mu = (\partial_0,\vec\nabla)$.

We release the state in Bunch-Davies vacuum at $t = 0$,
corresponding to conformal time $\eta =\eta_i\equiv -H^{-1}$.
Hence, the scale factor is normalized to $a=1$ when the state is
released so that $a>1$ throughout the evolution. Note that the
infinite future corresponds to $\eta \rightarrow 0^-$, so the
possible variation of causally related conformal coordinates in
either space or time is at most ${\Delta x} = {\Delta \eta} =
H^{-1}$. Applying the Schwinger-Keldish formalism \cite{JS,RJ} and
using dimensional regularization, the fully renormalized vacuum
expectation value (VEV) of the stress-energy tensor,
$\langle\Omega|T_{\mu\nu}(x)|\Omega\rangle$, is calculated
\cite{OW1,OW2} in this system. The energy density
$\rho=\langle\Omega|T_{00}|\Omega\rangle/a^2(\eta)$ and pressure
$p\delta_{ij}=\langle\Omega|T_{ij}|\Omega\rangle/a^2(\eta)$ are
obtained as
\begin{eqnarray}
\r_{\mbox{\tiny{ren}}} & = & \frac{\Lambda}{8\pi G} + \frac{\l
H^4}{2^6\pi^4} \Bigg\{ \frac12 \ln^2\left(a\right) + \frac29
a^{-3} - \frac12 \sum_{n=1}^{\infty}
\frac{n+2}{(n+1)^2} a^{-(n+1)} \Bigg\} + O(\lambda^2) \; , \label{rho} \\
p_{\mbox{\tiny{ren}}} & = & - \frac{\Lambda}{8\pi G} - \frac{\l
H^4}{2^6\pi^4} \Bigg\{\frac12 \ln^2\left(a\right) + \frac13
\ln\left(a\right) + \frac16 \sum_{n=1}^{\infty}
\frac{n^2-4}{(n+1)^2} a^{-(n+1)}\Bigg\} + O(\lambda^2) \; .
\label{pres}
\end{eqnarray}
Notice that $\rho_{\mbox{\tiny{ren}}}$ and $p_{\mbox{\tiny{ren}}}$
obey \cite{OW1,OW2} the covariant conservation law
$T_{\mu\nu}(x)\,^{;\nu}=0$, i.e.,
$\dot\rho_{\mbox{\tiny{ren}}}=-3H(\rho_{\mbox{\tiny{ren}}}
+p_{\mbox{\tiny{ren}}})$, where the dot denotes derivative with
respect to the comoving time $t$. Their sum, however, violates the
Weak Energy Condition (WEC) $\rho\!+\!p\!\geq\! 0$ on cosmological
scales \be \rho_{\mbox{\tiny{ren}}} + p_{\mbox{\tiny{ren}}} =
\frac{\l H^4}{2^6\pi^4} \Bigg\{-\frac13 \ln\left(a\right) +
\frac29 a^{-3} - \frac16 \sum_{n=1}^{\infty} \frac{n+2}{n+1}
a^{-(n+1)}\Bigg\} + O(\lambda^2) \; . \label{it} \ee Although the
value for $w+1$ is unobservably small in this model, the
calculation shows that quantum effects can induce a self-limiting
phase of super-acceleration in which a classically stable theory
violates the WEC on cosmological scales in the average of
$\rho\!+\!p$, not just in fluctuations about an average that obeys
the condition $\rho\!+\!p=0$. This is because inflationary
particle production causes the scalar to undergo a random walk
such that its average distance from the minimum of the potential
$\frac{\lambda}{4!}\varphi^4$ increases. In our model,
$\langle\Omega \vert \varphi^2(x) \vert \Omega\rangle\!=\! ({\rm
UV\; divergence})+H^2\ln(a)/{4\pi^2}+O(\lambda)$ \cite{VF}; recall
that $\ln(a)\!=\!Ht$. (See the calculations in
Sec.~\ref{sec:Stoch} for the $O(\lambda)$ and $O(\lambda^2)$
corrections.) Hence, after the ultraviolet divergence is removed,
the VEV of $\varphi^2$ gets pushed up its potential by
inflationary particle production. This increases the vacuum energy
which leads to the violation of the WEC by virtue of the covariant
conservation $\dot\rho\!=\!-3H(\rho+p)$; since $\dot\rho
>0$ due to inflationary particle production, $\rho+p$ has to be less than zero.

The process, however, must be self-limiting because (i) as the
scalar rises up its potential, the classical restoring force
$-\lambda\varphi^3/6$ pushes it back down, and (ii) the curvature
$\lambda\varphi^2/2$ associated with being away from the minimum
of the potential {\it acts like} a positive ``mass-squared'' to
reduce the inflationary particle production responsible for
pushing the scalar away from the configuration $\varphi\!=\!0$
where the potential is minimum. (In quantum field theory (QFT) the
mass-squared is calculated via self-energy diagrams as is
rigourously done in Ref.~\cite{BOW}. The VEV of the curvature of
the potential provides a heuristic picture to understand the
effect.) Since the classical restoring force (i) gets bigger as
the field rolls up its potential and the mass generation (ii) cuts
off particle production, the field cannot continue rolling up its
potential. It must eventually come to a halt. Indeed, Starobinsky
and Yokoyama showed \cite{SY} that $\langle\Omega \vert
\varphi^2(x) \vert \Omega\rangle$ asymptotes to the constant $3
H^2 \Gamma(3/4)/\pi\Gamma(1/4)\sqrt{\l}$ in this model, which {\it
proves} that the field strength does not grow forever. The
curvature of the potential, that acts like mass-squared, should
asymptote to $\lambda/2$ times this expectation value. They also
estimated the time scale for the process as $T \approx 18.7/ H
\sqrt{\l}$. Thus, by choosing $\lambda\ll 1$, it is possible to
have long duration for the effect. We assume $\lambda\ll 1$ in
this paper.

We study the stability of the system in this paper. To decide
whether the system is stable \cite{VF} or not, one needs to check
(i) if the VEV $\langle\Omega \vert \varphi^2(x) \vert
\Omega\rangle$ continues to grow without a bound, and (ii) if the
small, position-dependent perturbations grow. If neither happens,
the system is stable; otherwise, it is unstable. The above
arguments show that $\langle\Omega \vert \varphi^2(x) \vert
\Omega\rangle$ cannot continue to grow forever in the interacting
theory (it asymptotes to a constant). Checking criterion (ii) is
the main object of this paper. To do that, one solves the
quantum-corrected effective field equation at linearized order
\begin{equation} \Box \varphi(x) - \int d^4x'
M^2(x;x') \varphi(x') = 0 \; ,
\end{equation} and obtains the quantum-corrected mode function.
Although the scalar is classically massless in our model, quantum
processes generate a nonzero self-mass-squared $M^2(x;x')$.
Potential instabilities would come from the field developing a
negative mass-squared. In that case, the amplitude of the mode
function would be an increasing function of time, indicating
growth of perturbations and, hence, the instability. The fully
renormalized scalar self-mass-squared $M^2(x;x')$ is calculated
rigorously in Ref. \cite{BOW} at one- and two-loop order, using
the Schwinger-Keldish formalism. $M^2(x;x')$ is indeed {\it
positive} at one-loop. However, one must go to two-loop order to
see corrections of the derivative terms. To interpret the two-loop
result, and hence to check the stability of the system, one needs
to investigate how the self-mass-squared $M^2(x;x')$ modifies the
effective field equations and its solution, i.e., the
quantum-corrected mode function. If the amplitude of the solution
is a decreasing function of time, one can conclude that
perturbations do not grow; therefore, the model is stable.

The outline is as follows. In Sec.~\ref{sec:EffectiveMode}, we
define the effective mode equation, summarize the
Schwinger-Keldish formalism, and discuss our limitations in
solving the effective mode equation. In Sec.~\ref{sec:modefunc},
we solve the effective mode equation in late time limit and obtain
the mode function in leading logarithm approximation. Late time,
for us, means $\ln(a)\gg 1$. In Sec.~\ref{sec:Stoch}, we
alternatively compute the same mode function using Starobinsky's
stochastic inflation technique, and compare it with the result
obtained in Sec.~\ref{sec:modefunc}. Our conclusions are
summarized in Sec.~\ref{sec:Conc}.

\section{Effective Mode Equation for the MMC Scalar}
\label{sec:EffectiveMode}

In this section, we describe the operator formalism and effective
field equation correspondence. Then, we review the
Schwinger-Keldish formalism that one must use to calculate
expectation values. We use the one- and two-loop results
\cite{BOW} for the scalar self-mass-squared $M^2(x;x')$, obtained
by applying the Schwinger-Keldish formalism in our model, to write
down the effective (quantum-corrected) mode equation and discuss
how we ``solve'' it.

\subsection{Relation to Fundamental Operators}
\label{sec:2A} The relation between the fundamental Heisenberg
operator of scalar field $\varphi(x)$, and the $\mathbb C$-number
plane wave mode solution $\Phi(x;\vec{k})$ of the linearized
effective field equation can be given \cite{MW1,KW2} as
\begin{eqnarray}
\Phi(x;\vec{k})= \Bigl\langle \Psi_f \Bigl\vert
\Bigl[\varphi(x),\alpha^{ \dagger}(\vec{k})\Bigr] \Bigr\vert
\Psi_i\Bigr\rangle \; . \label{calphi}
\end{eqnarray}
Here $\vert \Psi_i\rangle$ and $\vert \Psi_f\rangle$ are the
states, and $\alpha^{\dagger}( \vec{k})$ is the free creation
operator. In flat space scattering problems, $\vert \Psi_i\rangle$
and $\vert \Psi_f\rangle$ correspond to the states whose wave
functionals are free vacuum in the asymptotic past and future,
respectively. The universe, however, begins at a finite time, and
evolves to some unknown state in the asymptotic future. Therefore,
in cosmology, we release the universe from a prepared state at a
given time and then let it evolve. We seek to know expectation
values in the presence of this state. This corresponds to the
choice $\vert \Psi_f \rangle = \vert \Psi_i \rangle$. For
computational convenience, we assume that both of the states are
free vacuum at $\eta \!=\! \eta_i$. Had the choices of flat space
scattering theory been used, acausal effective field equations
would have been obtained. The matrix elements of Hermitian
operators would also be complex in that case.

To define the free creation and annihilation operators, recall
that the full Lagrangian density $\mathcal{L}$ of the MMC scalar
is \be {\cal L} = -\frac{(1+\delta Z)}{2}\sqrt{-g}g^{\mu\nu}
\partial_{\mu} \varphi \partial_{\nu} \varphi -
\frac{(\lambda +\delta\lambda)}{4!}\sqrt{-g}\varphi^4-\frac{\delta
m^2}{2} \sqrt{-g}\varphi^2 \; . \label{fulLag}\ee The field
strength ($\delta Z$), coupling constant ($\delta \lambda$), and
mass ($\delta m^2$) counterterms are needed to remove divergences
at one- and two-loop order in the scalar self-mass squared. It
turns out that $\delta Z$ and $\delta\lambda$ are of order
$\lambda^2$, whereas $\delta m^2$ has contributions of order
$\lambda$ and $\lambda^2$ \cite{BOW}.

Let us now integrate the invariant field equation of the MMC
scalar
\begin{eqnarray}
\dd_{\mu} \Bigl(\sqrt{-g} g^{\mu\nu} \dd_{\nu} \varphi\Bigr)
-\frac{\sqrt{-g}}{1 \!+\! \delta Z}\left[\frac{(\lambda+\delta
\lambda)}{6} \varphi^3 + \delta m^2\varphi\right] = 0 \; .
\label{phieqn}
\end{eqnarray}The result is
\begin{eqnarray}
\varphi(x) & = & \varphi_0(x) + \int_{\eta_i}^0 d\eta' \int
d^{D-1}x' G(x;x') I[\varphi(x')] \; , \label{phi0}
\end{eqnarray}
where $\varphi_0(x)$ is the free field. We define the interaction
term as
\begin{eqnarray}
I[\varphi]\equiv\frac{\sqrt{-g}}{1 \!+\! \delta
Z}\left[\frac{(\lambda+\delta \lambda)}{6} \varphi^3 + \delta
m^2\varphi\right]  \; .
\end{eqnarray}
The Green's function $G(x;x')$ is any solution of the equation
\begin{equation}
\partial_{\mu} \Bigl(\sqrt{-g} g^{\mu\nu} \partial_{\nu} G(x;x') \Bigr) =
\delta^D(x \!-\! x') \; .
\end{equation}
Although the Green's functions would obey Feynman boundary
conditions for flat space scattering problems, it is more natural
to use retarded boundary conditions in cosmology. The fundamental
field operator $\varphi(x)$, on the other hand, is unique. It does
not depend on the choices of the boundary conditions for the
Green's functions or on $\eta_i$. What changes with those choices
is the free scalar $\varphi_0(x)$. Because $\varphi_0(x)$ obeys
the linearized equations of motion and agrees with the full fields
at $\eta \!=\! \eta_i$, it can be expanded in terms of free
creation and annihilation operators $\alpha(\vec{k})$ and
$\a^{\dagger}(\vec{k})$ as
\begin{eqnarray}
\varphi_0(x)= \int\frac{d^{D-1}k}{(2\pi)^{D-1}} \Bigl\{u(\eta,k)
e^{i \vec{k} \cdot \vec{x}} \alpha(\vec{k}) + u^*(\eta,k)
e^{-i\vec{k} \cdot \vec{x}} \a^{\dagger}(\vec{k}) \Bigr\} \;
,\label{varphizero}
\end{eqnarray}
where the Bunch-Davies mode function \cite{BD} \beeq
u(\eta,k)=\frac{H}{\sqrt{2 k^3}} \Bigl(1 + i k \eta\Bigr) e^{-i k
\eta} \; . \label{u}\eneq Although the creation and annihilation
operators change as different Green's functions are used in
Eq.~(\ref{phi0}), their nonzero commutation relation remains fixed
\begin{eqnarray}
[\alpha(\vec{k}),\alpha^{\dagger}(\vec{k}')] & = & (2\pi)^{D-1}
\delta^{D-1}(\vec{k} \!-\! \vec{k}')\; .\label{commut}
\end{eqnarray}
By iterating Eq.~(\ref{phi0}), one can expand the full field
$\varphi(x)$ in terms of the free field $\varphi_0(x)$ as
\begin{eqnarray}
\varphi(x) & = & \varphi_0(x) + \int_{\eta_i}^0 d\eta' \int
d^{D-1}x'
G_{\rm ret}(x;x') I[\varphi(x')]  \\
& = & \varphi_0(x) + \int_{\eta_i}^0 d\eta' \int d^{D-1}x' G_{\rm
ret}(x;x') I[\varphi_0(x')] + \ldots \; .
\end{eqnarray}
Hence, choosing $\vert \Psi_f\rangle \!=\! \vert \Psi_i\rangle$ as
free vacuum at $\eta_i$, one can see \cite{MW1,KW2} that the
quantum-corrected plane wave mode solution~(\ref{calphi}) yields
\begin{equation}
\Phi(x;\vec{k}) = \Bigl\langle \Omega \Bigl\vert \Bigl[\varphi(x),
\alpha^{\dagger}(\vec{k})\Bigr] \Bigr\vert \Omega \Bigr\rangle =
u(\eta,k) e^{i\vec{k} \cdot \vec{x}} + O(\l) \; . \label{effmod}
\end{equation}
The $O(\l)$ and $O(\l^2)$ corrections in Eq.~(\ref{effmod}) are
obtained in Sec.~\ref{sec:modefunc}, by ``solving'' the
quantum-corrected effective field equation at one- and two-loop
order. In Sec.~\ref{sec:Stoch}, we obtain the same corrections by
calculating commutator~(\ref{calphi}) stochastically. The results
yielded by the two approaches are in perfect agreement.
\subsection{Schwinger-Keldish Formalism}
\label{sec:S-K} Because of the fact that ``in'' $(t\rightarrow
-\infty)$ vacuum is not equal to the ``out'' $(t\rightarrow
\infty)$ vacuum in de Sitter background, we need to calculate
expectation values, rather than in-out matrix elements. This is
done by applying the Schwinger-Keldish formalism~\cite{JS,RJ}. The
endpoints of propagators acquire a $\pm$ polarity, in this
formalism. Hence, every propagator $i\Delta(x;x')$ of the in-out
formalism generalizes to four Schwinger-Keldysh propagators:
$i\Delta_{\scriptscriptstyle ++}(x;x')$,
$i\Delta_{\scriptscriptstyle +-}(x;x')$, $i\Delta_{
\scriptscriptstyle -+}(x;x')$ and $i\Delta_{\scriptscriptstyle
--}(x;x')$. Each propagator can be obtained from the Feynman
propagator by replacing the de Sitter conformal coordinate
interval
\begin{equation}
\D x^2(x,x')=\Delta x^2_{\scriptscriptstyle ++}(x;x') \equiv
\Bigl\Vert \vec{x} \!-\! \vec{x}' \Bigr\Vert^2 - \Bigl(\vert \eta
\!-\! \eta'\vert \!-\! i \delta \Bigr)^2\label{plusplus}
\end{equation}
with the appropriate coordinate interval,\begin{eqnarray} \Delta
x^2_{\scriptscriptstyle +-}(x;x') & \equiv & \Bigl\Vert \vec{x}
\!-\!
\vec{x}' \Bigr\Vert^2 - \Bigl(\eta \!-\! \eta' \!+\! i \delta \Bigr)^2
=(\Delta x^2_{\scriptscriptstyle -+}(x;x'))^* \; , \label{plusminus}\\
\Delta x^2_{\scriptscriptstyle --}(x;x') & = & (\Delta
x^2_{\scriptscriptstyle ++}(x;x'))^*\; .
\end{eqnarray}
Vertices are either all $+$ or all $-$. A $+$ vertex is the usual
one of the in-out formalism, whereas the $-$ vertex is its
conjugate.

Because each external line can be either $+$ or $-$ in the
Schwinger-Keldysh formalism, each $N$-point 1PI function of the
in-out formalism corresponds to $2^N$ Schwinger-Keldysh $N$-point
1PI functions. The Schwinger-Keldysh effective action is the
generating functional of these 1PI functions, so it depends upon
two background fields $\varphi_+(x)$ and $\varphi_-(x)$. For
example, there are four Schwinger-Keldysh 2-point 1PI functions
$M^2_{\scriptscriptstyle \pm\pm}(x;x')$. The $++$ one is the same
as the in-out self-mass-squared and the others are related as the
propagators
\begin{equation}
-i M^2_{\scriptscriptstyle --}(x;x') = \Bigl(-i
M^{2}_{\scriptscriptstyle ++}( x;x') \Bigr)^* \; , \qquad -i
M^2_{\scriptscriptstyle -+}(x;x') = \Bigl( -i
M^2_{\scriptscriptstyle +-}(x;x')\Bigr)^* .
\end{equation}
The various self-mass-squared terms enter \cite{RJ} the effective
action as follows:
\begin{eqnarray}
\lefteqn{\Gamma[\varphi_+,\varphi_-] = S[\varphi_+] - S[\varphi_-]} \nonumber \\
& & - \frac12 \int d^Dx \int d^Dx' \left\{ \matrix{ \varphi_+(x)
M^2_{ \scriptscriptstyle ++}(x;x') \varphi_+(x') + \varphi_+(x)
M^2_{\scriptscriptstyle +-}(x;x') \varphi_-(x') \cr +\varphi_-(x)
M^2_{\scriptscriptstyle -+}(x;x') \varphi_+(x') + \varphi_-(x)
M^2_{\scriptscriptstyle --}(x;x') \varphi_-(x') } \right\} +
O(\varphi^3_{\pm}) , \qquad
\end{eqnarray}
where $S[\varphi]$ is the classical scalar action. The effective
field equations of the Schwinger-Keldysh formalism are obtained by
varying with respect to either polarity, and then setting the two
polarities equal \cite{RJ}. Up to order $O(\varphi^2)$, we have
\begin{eqnarray}
\frac{\delta \Gamma[\varphi_{ \scriptscriptstyle\pm}]}{\delta
\varphi_{\scriptscriptstyle +}(x)}
\Biggl\vert_{\varphi_{\scriptscriptstyle \pm} = \varphi}
\!\!\!\!\!\!\!\!\!=
\partial_{\mu} \Bigl(\sqrt{-g}
g^{\mu\nu} \partial_{\nu} \varphi(x)\Bigr)\!-\!\int_{\eta_i}^0\!\!
d\eta'\!\int\!d^3x' \!\Bigl\{M^2_{ \scriptscriptstyle ++}(x;x') +
M^2_{\scriptscriptstyle +-}(x;x')\Bigr\} \varphi(x')\; .
\end{eqnarray} Note that
we have taken the regularization parameter $D$ to its unregulated
value of $D \!=\! 4$, in view of the fact that the
self-mass-squared is assumed to be fully renormalized.  It is this
{\it linearized} effective field equation which $\Phi(x;\vec{k})$
(Eq.~(\ref{calphi})) obeys \cite{MW1,KW2}
\begin{equation}
\dd_\mu\Bigl(\sqrt{-g} g^{\mu\nu} \partial_{\nu}
\Phi(x;\vec{k})\Bigr) - \int_{\eta_i}^0 d\eta' \int d^3x'
\Bigl\{M^2_{\scriptscriptstyle ++}(x;x') + M^2_{\scriptscriptstyle
+-}(x;x') \Bigr\} \Phi(x';\vec{k}) = 0 \; . \label{lineqn}
\end{equation} Thus, the two renormalized 1PI 2-point functions we need are
$M^2_{\scriptscriptstyle ++}(x;x')$ and $M^2_{\scriptscriptstyle
+-}(x;x')$. At one-loop order, we have \cite{BOW}\be
M^2_{\scriptscriptstyle 1++}(x;x') = \frac{\l H^2}{8\pi^2} \,a^4
\ln(a) \delta^4(x-x')+O(\l^2)\; .\label{1loop}\ee The $+-$ case
vanishes at this order because there is no mixed interaction.
Fully renormalized two-loop results for the $++$ and $+-$ cases
are \cite{BOW} \be & & \!\!\!\!\!\!M^2_{\mbox{{\tiny 2++}}}=
\frac{i \l^2}{2^9 \pi^6} \Bigg\{\frac{a a'}{24} \dd^4
\Biggl[\frac{\ln\left(\mu^2 \D x^2_{\scriptscriptstyle
++}\right)}{\D x^2_{\scriptscriptstyle ++}} \Biggr] \!-\! H^2 (a
a')^2 \dd^2 \Biggl[ \ln\Bigl(\frac{H e^{\frac34}}{2 \mu}\!\Bigr)
\frac{\ln\left(\mu^2 \D x^2_{ \scriptscriptstyle ++}\right)}{\D
x^2_{\scriptscriptstyle ++}} \!+\! \frac{\ln^2\left(\mu^2 \D
x^2_{\scriptscriptstyle ++}\right)}{4 \D x^2_{
\scriptscriptstyle ++}}\Biggr] \nonumber \\
& & - H^4 (a a')^3 \frac{ \ln^2\left(\frac{\sqrt{e}}4 H^2 \D
x^2_{\scriptscriptstyle ++}\right)}{\D x^2_{\scriptscriptstyle
++}} \!+\! \frac{H^6}6 (a a')^4 \! \ln^3\Bigl(\frac{\sqrt{e}}4 H^2
\D x^2_{
\scriptscriptstyle ++} \Bigr) \Bigg\} \nonumber \\
& & + \frac{\l^2}{2^9 3\, \pi^4} \, a^2 \Biggl\{-\ln(a) \dd^2
\!+\! \Bigl(2 \ln(a) \!+\! 1\Bigr) H a \partial_0 \Biggr\}
\delta^4(x \!-\! x')
\nonumber \\
& & + \frac{\lambda^2 H^2}{2^7 \pi^4} \Biggl\{-\frac49 \ln^3(a)
\!-\! \frac{23}{18} \ln^2(a) \!+\! \Biggl[\frac{13}3 \!+\! 3
\ln\Bigl(\frac{H}{2 \mu} \Bigr) \!-\! \frac29 \pi^2 \Biggr] \ln(a)
\Biggl\} a^4 \delta^4(x \!-\! x')
\nonumber \\
& & + \frac{\lambda^2 H^2}{2^7 \pi^4} \Biggl\{ \frac{a^{-3}}{81}
\!-\! \sum_{n=1}^\infty \frac{n+5}{(n+1)^3}a^{-(n+1)} \!+\! 4
\sum_{n=1}^\infty \frac{a^{-(n+2)}}{(n+2)^3} \!+\! 4
\sum_{n=1}^\infty\frac{a^{-(n+3)}}{n(n+3)^3} \Biggr\} a^4
\delta^4(x \!-\! x') \; . \label{M2++}\\
& & \!\!\!\!\!\!M^2_{\mbox{{\tiny 2$+-$}}}= -\frac{i \l^2}{2^9
\pi^6} \Bigg\{\frac{a a'}{24} \dd^4 \Biggl[\frac{\ln\left(\mu^2 \D
x^2_{\scriptscriptstyle +-}\right)}{\D x^2_{\scriptscriptstyle
+-}} \Biggr] \!-\! H^2 (a a')^2 \dd^2 \Biggl[ \ln\Bigl(\frac{H
e^{\frac34}}{2 \mu}\!\Bigr) \frac{\ln\left(\mu^2 \D x^2_{
\scriptscriptstyle +-}\right)}{\D x^2_{\scriptscriptstyle +-}}
\!+\! \frac{\ln^2\left(\mu^2 \D x^2_{\scriptscriptstyle
+-}\right)}{4 \D x^2_{
\scriptscriptstyle +-}}\Biggr] \nonumber \\
& & - H^4 (a a')^3 \frac{ \ln^2\left(\frac{\sqrt{e}}4 H^2 \D
x^2_{\scriptscriptstyle +-}\right)}{\D x^2_{\scriptscriptstyle
+-}} \!+\! \frac{H^6}6 (a a')^4 \! \ln^3\Bigl(\frac{\sqrt{e}}4 H^2
\D x^2_{ \scriptscriptstyle +-} \Bigr) \Bigg\} \; .\label{M2+-}\ee
The $++$ and $+-$ terms in (\ref{lineqn}) exactly cancel for
$\eta' \!>\! \eta$ and also, in the limit $\delta \!\rightarrow\!
0$, for $x^{\prime \mu}$ outside the light-cone of $x^{\mu}$. This
is how the Schwinger-Keldysh formalism gives causal effective
field equations. In the next section, we discuss what we mean by
``solving'' the quantum-corrected effective mode equation
(\ref{lineqn}). The one- and two-loop corrected mode solution is
obtained in late time limit, i.e., for $\ln(a)\gg 1$, in
Sec.~\ref{sec:modefunc}.

\subsection{Solving the quantum-corrected effective mode equation}
\label{sec:solve} Here we discuss the limitations that one has in
solving the effective mode equation (\ref{lineqn}). The full
scalar self-mass-squared can be expressed, as a series, in powers
of the loop counting parameter $\l$
\begin{equation}
M^2_{\scriptscriptstyle ++}(x;x') + M^2_{\scriptscriptstyle
+-}(x;x') = \sum_{\ell=1}^{\infty} \l^{\ell}
\mathcal{M}^2_{\ell}(x;x') \; .\label{mass}
\end{equation}
The first limitation is that we have only the $\ell \!=\! 1$ and
$\ell \!=\! 2$ terms \be
M^2_{\scriptscriptstyle 1++}(x;x')\;&=&\l\, \mathcal{M}^2_1(x;x') \; , \label{1el}\\
M^2_{\scriptscriptstyle 2 ++}(x;x') + M^2_{\scriptscriptstyle 2
+-}(x;x')\;&=& \l^2 \,\mathcal{M}^2_2(x;x') \; ,\label{2el}\ee
which are given by Eqs.~(\ref{1loop}), (\ref{M2++}) and
(\ref{M2+-}), respectively. So we can only solve the effective
mode equation perturbatively. We first substitute a series
solution of the form
\begin{equation}
\Phi(x;\vec{k}) \equiv u(\eta,k) e^{i\vec{k} \cdot \vec{x}} +
\sum_{\ell=1}^{ \infty} \l^{\ell} \Phi_{\ell}(\eta,k) e^{i \vec{k}
\cdot\vec{x}} \label{pert}
\end{equation}
into Eq.~(\ref{lineqn}) and then solve the equation order by order
in powers of $\l$ and $\l^2$. The zeroth order ($\ell=0$) solution
of $\Phi_{\ell}$ is the well-known Bunch-Davies mode function
$u(\eta, k)$ (Eq.~(\ref{u})) times the exponential $e^{i\vec{k}
\cdot \vec{x}}$.

The second limitation is due to the lower bound ``$\eta_i$'' on
the temporal integration in Eq.~(\ref{lineqn}). We release the
universe in free vacuum at time $\eta \!=\! \eta_i$. Little is
known about the wave functionals of interacting QFTs in curved
space, but free vacuum can hardly be realistic. In fact, all of
the finite energy states of interacting flat space QFTs have
important corrections. Similar corrections are expected in curved
space, too. Although it is possible to correct the free state
functionals perturbatively as in nonrelativistic quantum
mechanics, the usual procedure in flat space QFT is to release the
system in free vacuum at asymptotic past, and let the infinite
time evolution resolve the difference between free vacuum and true
vacuum into shifts of the mass, field strength and background
field \cite{BjD}. In cosmology, however, one cannot typically
apply this procedure, for the reasons noted in Sec.~\ref{sec:2A}.
One can still correct the state wave functionals perturbatively,
though. Corrections to the initial state would appear as new
interaction vertices on the initial value surface. They are
expected to have a large effect on the expectation values of
operators near the initial value which would decay in the late
time limit. For example, it is the exponentially falling portions
of the renormalized stress-energy tensor (\ref{rho}) and
(\ref{pres})
\begin{eqnarray}
\rho_{\mbox{\tiny{falling}}}& = & \frac{\lambda H^4}{2^6 \pi^4}
\Biggl\{ \frac{2}{9} a^{-3} \!-\! \frac12 \sum_{n=1}^{\infty}
\frac{n\!+\!2}{(n\!+\!1)^2}a^{-(n+1)}
\Biggr\} + O(\lambda^2) \; , \label{Drho} \\
p_{\mbox{\tiny{falling}}}& = & \frac{\lambda H^4}{2^6 \pi^4}
\Biggl\{- \frac1{6} \sum_{n=1}^{\infty}
\frac{n^2\!-\!4}{(n\!+\!1)^2} a^{-(n+1)} \Biggr\} + O(\lambda^2)
\label{Dp}
\end{eqnarray}
that it is conjectured \cite{OW2} can be absorbed into an order
$\lambda$ correction of the initial ($a\!=\!1$) free Bunch-Davies
vacuum state. The fact that they fall off as one evolves away from
the initial value surface suggests that they can be absorbed into
a kind of local interaction there, leaving only the infrared
logarithms\begin{eqnarray} \r_{\mbox{\tiny{conj}}} & = &
\frac{\Lambda}{8\pi G} + \frac{\l H^4}{2^6\pi^4}
\Bigg\{ \frac12 \ln^2\left(a\right) \Bigg\} + O(\lambda^2) \; , \\
p_{\mbox{\tiny{conj}}} & = & - \frac{\Lambda}{8\pi G} - \frac{\l
H^4}{2^6\pi^4} \Bigg\{\frac12 \ln^2\left(a\right) + \frac13
\ln\left(a\right)\Bigg\} + O(\lambda^2) \; .
\end{eqnarray}
Notice that they are separately conserved, i.e.,
$\dot\rho_{\mbox{\tiny{conj}}}\!\!=\!\!-3H(\rho_{\mbox{\tiny{conj}}}+p_{\mbox{\tiny{conj}}})$
and
$\dot\rho_{\mbox{\tiny{falling}}}\!\!=\!\!-3H(\rho_{\mbox{\tiny{falling}}}+p_{\mbox{\tiny{falling}}})$.
This is exactly what would be the case if they could be cancelled
by a new interaction vertex. Note that Eqs. (\ref{Drho}) and
(\ref{Dp}) diverge on the initial value surface at $a\!=\!1$ which
indicates that free vacuum is very far away from any physically
accessible state. Thus, although Eq.~(\ref{lineqn}) determines the
quantum corrections to the mode function (\ref{effmod}) for free
vacuum, that mode function has little physical relevance, because
free vacuum is inaccessible. To find physically relevant mode
functions, which are also valid for initial times, the corrections
to the state wave functional must be included. Unfortunately, we
have neither order $\lambda$ nor order $\l^2$ corrections to the
state wave functional. It therefore makes no sense to solve
Eq.~(\ref{lineqn}) for all times. The effects of the state
corrections, however, must fall off at late times ($\ln(a)\gg 1$)
as in Eqs.~(\ref{Drho}) and (\ref{Dp}). Because of time evolution,
initially free vacuum and true vacuum become indistinguishable, as
in flat space QFT~\cite{BjD}. Hence, we may obtain valid
information from Eq.~(\ref{lineqn}) by solving it in late time
limit. That is the subject of the next section.

\section{Effective Mode Function for the MMC Scalar}
\label{sec:modefunc}

The linearized effective field equation that the MMC scalar mode
solution $\Phi(x, \vec{k})$ obeys is given in Eq.~(\ref{lineqn}).
Using Eq.~(\ref{mass}), (\ref{1el}) and (\ref{pert}), one obtains
the integro-differential equation for the one-loop correction
$\Phi_1(\eta, \vec{k})$ to the classical mode function $u(\eta,
k)$ \be &&a^2[\dd_0^2\!+\!2Ha\dd_0\!+\!k^2]\Phi_1(\eta,
k)=-\!\int_{\eta_i}^0 d\eta'\int d^3x'
\mathcal{M}^2_1(x;x')u(\eta', k)e^{-i \vec{k}
\cdot(\vec{x}-\vec{x}\,')}\nonumber\\
&&=-\!\int_{\eta_i}^0 d\eta'\int d^3x' \frac{H^2}{8\pi^2} \,a^4
\ln(a)\delta^4(x\!-\!x')u(\eta',k) e^{-i \vec{k}
\cdot(\vec{x}-\vec{x}\,')}=-\frac{H^2}{8\pi^2} u(\eta,k)\,a^4
\ln(a)\; .\hskip 1cm\label{order1}\ee As is discussed in
Sec.~\ref{sec:solve}, the only sensible and physically interesting
regime in which we can solve the effective mode equation is the
late time limit $\ln(a)\gg 1$. The zeroth order mode function
$u(\eta, k)$ can be replaced by its limit $u(0,k)=H/\sqrt{2k^3}$
in this regime. Solving Eq.~(\ref{order1}) in late time limit, we
find \beeq\Phi_1(\eta, k)\sim-\frac{1}{2^4
3\pi^2}u(0,k)\left\{\ln^2(a)
-\frac{2}{3}\ln(a)\right\}\label{Phi1}\; ,\eneq in leading
logarithm orders.

The order $\l^2$ correction $\Phi_2(\eta, k)$, on the other hand,
has contributions due to both one- and two-loop self-mass-squared
terms. It obeys \be &&a^2[\dd_0^2+2Ha\dd_0+k^2]\Phi_2(\eta,
k)\nonumber\\&&=-\int_{\eta_i}^0 d\eta'\int
d^3x'\left\{\mathcal{M}_1^2(x;x')\Phi_1(\eta',
k)+\mathcal{M}_2^2(x;x')u(\eta', k)\right\}e^{-i \vec{k}
\cdot(\vec{x}- \vec{x}\,')}\; .\label{IntegroPhi2}\ee The first
integral is evaluated, in leading logarithm order, by inserting
Eqs.~(\ref{1loop}) and (\ref{Phi1}) into Eq.~(\ref{IntegroPhi2}).
We find \beeq -\int_{\eta_i}^0 d\eta'\int d^3x'
\mathcal{M}_1^2(x;x')\Phi_1(\eta', k)\sim\frac{H^2
}{2^7\,3\pi^4}u(0,k)a^4\left[\ln^3(a)-\frac{2}{3}\ln^2(a)\right]\;
. \label{M1Phi1}\eneq The second integral is evaluated, in late
time limit, in Appendix~\ref{App:secondint}. Expanding in terms of
powers of infrared logarithms, we find\be
&&\!\!\!\!\!\!-\int_{\eta_i}^0 d\eta'\int d^3x'
\mathcal{M}_2^2(x;x')u(\eta', k)e^{-i \vec{k} \cdot(\vec{x}-
\vec{x}\,')}\longrightarrow -u(0, k)\int_{\eta_i}^0 d\eta'\int
d^3x'\mathcal{M}_2^2(x;x')\nonumber\\
&&\!\!\!\!\!\!\sim\frac{H^2
}{2^4\,3^2\pi^4}u(0,k)a^4\left[\ln^3(a)+\frac{23}{16}\ln^2(a)
+\left(\frac{27}{8}\ln\Bigl(\frac{2\mu}{H}\Bigr)
-\frac{189}{32}+\frac{\pi^2}{2}\right)\ln(a)\right]\;
.\label{M2u}\ee Using Eqs.~(\ref{M1Phi1}) and (\ref{M2u}) in
Eq.~(\ref{IntegroPhi2}) yields \be
&&a^2[\dd_0^2+2Ha\dd_0+k^2]\Phi_2(\eta,
k)\nonumber\\&&\rightarrow\frac{H^2 }{2^7
3^2\pi^4}u(0,k)a^4\left[11\ln^3(a)+\frac{19}{2}\ln^2(a)
+\left(27\ln\Bigl(\frac{2\mu}{H}\Bigr)
-\frac{189}{4}+4\pi^2\right)\ln(a)\right]\; .\ee In leading
orders, the solution for $\Phi_2(\eta, k)$ is
\be\!\!\!\!\!\Phi_2(\eta, k)\!\sim\!\frac{1}{2^8
3^3\pi^4}u(0,k)\Bigg\{\!\!\frac{11}{2}\ln^4(a)
\!-\!\ln^3(a)\!+\!\!\left[27\ln\Bigl(\frac{2\mu}{H}\Bigr)
\!\!-\!\frac{185}{4}\!+\!4\pi^2\right]\!\!\left[\ln^2(a)
\!-\!\frac{2}{3}\ln(a)\right]\!\!\Bigg\}\label{Phi2}\; .\ee Thus,
keeping the leading logarithm terms at each order of perturbation
(i.e., in $\Phi_1$ and $\Phi_2$), we find that the
quantum-corrected mode solution (\ref{pert}) asymptotes to \be
&&\Phi(x, \vec{k})\sim u(0,k) e^{i \vec{k}
\cdot\vec{x}}\Bigg\{1-\frac{1}{2^4
3\pi^2}\lambda\ln^2(a)+\frac{11}{2^9 3^3\pi^4}\l^2\ln^4(a)
\Bigg\}+O(\l^3)\; .\label{Modefunc}\ee One can immediately see
from Eq.~(\ref{Modefunc}) that perturbation theory breaks down
when $\ln{\left(a(t)\right)}$ is of order $1/\sqrt{\lambda}$.
This, however, does not invalidate the reliability of our late
time $(\ln(a)\!\gg\!1)$ solution, because by choosing $\lambda\ll
1$, as we assume in this paper, one can have a long period of time
during which $1\ll\ln(a)\ll 1/\sqrt{\lambda}$.

Equation~(\ref{Modefunc}) also shows that, at $t=0$, the mode
solution $\Phi(x, \vec{k})$ is equal to the well-known
(Bunch-Davies) classical result $u(0, k)
e^{i\vec{k}\cdot\vec{x}}$. As time goes on, it decreases
proportional to the factor $1-\lambda H^2
t^2/48\pi^2+O(\lambda^2)$ (the stochastic calculation of
Sec.~\ref{sec:Stoch} yields the same result). Thus, the amplitude
of quantum-corrected mode function (hence, of the field) is
reduced (consistent with the model developing a {\it positive}
mass-squared, as one- and two-loop self-mass-squared terms and the
VEV of the curvature of the potential imply; see the discussion in
the next section). This means that linear perturbations do not
grow in this system; therefore, it is stable.

\section{Stochastic Analysis}

\label{sec:Stoch}Starobinsky developed a stochastic inflation
technique \cite{Staro,SY} which gives the leading infrared
logarithms at each order in perturbation theory. Recently, his
technique was proven to all orders and extended to various models
\cite{W2}. In this section, we introduce the stochastic technique
briefly, and use it to calculate the quantum-corrected plane wave
mode solution~(\ref{calphi}) and the VEV of the curvature
(associated with the field being away from the minimum) of the
potential which acts like a mass-squared in the classical action.

The equation of motion for the scalar field with quartic
self-interaction in $D=3+1$ dimensional de Sitter background is
\begin{equation}
\ddot{\varphi}(t,\vec{x}) +  3H \dot{\varphi}(t,\vec{x}) -
\frac{\nabla^2}{a^2} \varphi(t,\vec{x}) + \frac{\lambda}{6}
\varphi^3(t,\vec{x}) \; = \; 0 \;\; . \label{feq}
\end{equation}
The solution of Eq.~(\ref{feq}) can be obtained by iterating
\begin{eqnarray}
\varphi(t,\vec{x}) & = & \varphi_0(t,\vec{x})
-\frac{\lambda}{6}\int_{0}^t dt' a^3(t')\int d^3x' G_{\rm ret}(t,
\vec{x};t', \vec{x}') \varphi^3(t',\vec{x}')  \; ,
\end{eqnarray}
where the retarded Green's function \cite{W2} is \be G_{\rm
ret}\!\equiv\!\frac{H^2}{4\pi}
\Theta(t\!-\!t')\Bigl[\frac{\delta(H\|\vec{x}\!-\!\vec{x}'\|+\!a^{-1}(t)\!-\!a^{-1}(t'))}
{a(t)a(t')H\|\vec{x}\!-\!\vec{x}'\|}\!+\!\Theta(H\|\vec{x}\!-\!\vec{x}'
\|\!+\!a^{-1}(t)\!-\!a^{-1}(t'))\Bigr]\; .\ee As in
Eq.~(\ref{varphizero}), the free field $\varphi_0(t, \vec{x})$ can
be expanded in terms of mode function $u(t, k)$ and annihilation
and creation operators $\alpha(\vec{k})$ and
$\alpha^\dagger(\vec{k})$, satisfying the canonical commutation
relation (\ref{commut}) in $D\!=\!3+\!1$ dimensions. Starobinsky's
stochastic technique cuts out the ultraviolet modes $k>Ha$ of the
field and applies the following rules to the equation of motion:
(i) retain only the term with the smallest number of derivatives
of the field, (ii) replace the field variable by a stochastic
variable, and (iii) subtract the stochastic source term $f$ for
each time derivative of the field. Applying these rules to
Eq.~(\ref{feq}) yields
\begin{equation}
\ddot{\varphi}+3 H \dot{\varphi} - \frac{\nabla^2}{a^2}\varphi
+\frac{\lambda}{6} \varphi^3=0 \longrightarrow  3 H
\dot{\varphi}+\frac{\lambda}{6} \, \varphi^3 =0 \longrightarrow 3
H \left(\dot{\phi}-f_{\phi}\right)+\frac{\lambda}{6}\phi^3=0 \;
.\label{rueq}
\end{equation} (The scalar field $\varphi$ became a stochastic
field $\phi$.) The source term $f_\phi$ is the time derivative of
the {\it infrared truncated} free field (\ref{varphizero})
\begin{equation}
\phi_0(t,\vec{x})\equiv\int \frac{d^3k}{(2\pi)^3} \; \theta( Ha(t)
- k) \, \frac{H}{\sqrt{2 k^3}} \, \Biggl\{ \, e^{i \vec{k} \cdot
\vec{x}} \, \alpha(\vec{k}) \, + \, e^{-i \vec{k} \cdot \vec{x}}
\, \alpha^{\dagger}(\vec{k}) \, \Biggr\} \; . \label{phizero}
\end{equation}
Here the leading infrared limit of the Bunch-Davies mode
function~(\ref{u}) $u(t, k)\sim H/\sqrt{2k^3}$ is used. Hence, \be
f_{\phi}(t,\vec{x})\equiv\dot{\phi}_0(t,\vec{x}) & = & \int
\frac{d^3k}{(2\pi)^3} \; \delta( k \!-\! H a(t)) \;
\frac{H^2}{\sqrt{2k}} \; \Biggl\{ \, e^{i \vec{k} \cdot \vec{x}}
\, \alpha(\vec{k}) \, + \, e^{-i \vec{k} \cdot \vec{x}} \,
\alpha^{\dagger}(\vec{k}) \, \Biggr\} \; . \label{stoksour} \ee In
Eq.~(\ref{rueq}), we obtained a Langevin-like equation which can
be recast as
\begin{equation}
\dot{\phi}(t,\vec{x}) = f_{\phi}(t,\vec{x})-\frac{\l}{18H}
\phi^3(t,\vec{x}) \; . \label{stokeq}
\end{equation}
In this section, we use this equation to stochastically calculate
(i) the VEV of the curvature associated with being away from the
minimum of the potential, i.e., $\lambda\langle \Omega \vert
\phi^2(x)\vert\Omega\rangle/2$, which acts like field-dependent
mass-squared, and (ii) the quantum-corrected mode
function~(\ref{calphi}), i.e., $\langle\Omega\vert[\phi(x),
\alpha^{\dagger}(\vec{k})]\vert \Omega\rangle$. As a check, we
calculate (i) also using perturbative QFT at one- and two-loop
order and show that the two realizations agree perfectly in
leading logarithm order. The quantum-corrected mode function was
already obtained in Sec.~\ref{sec:modefunc} by applying QFT.
Comparing Eq.~(\ref{Modefunc}) of Sec.~\ref{sec:modefunc} with the
stochastic result for (ii) will show that the agreement is again
perfect in leading logarithm order.

In stochastic calculations (i) and (ii), we express $\phi(x)$ in
terms of the infrared truncated free field $\phi_0$
perturbatively, by first integrating Eq.~(\ref{stokeq}) and then
iterating the result up to the desired power of $\lambda$:\be
\phi(t, \vec{x})\!\!&=&\!\!\phi_0(t, \vec{x})\!-\!\frac{\l}{18H}
\int_0^t
dt'\phi^3(t',\vec{x}) \label{LIReqn}\hspace{9.5cm} \nonumber\\
&=&\!\!\phi_0(t, \vec{x})\!-\!\frac{\l}{18H}\int_0^t\!\!dt'
\phi^3_0(t', \vec{x})\!+\!\frac{\l^2}{2^2 3^3 H^2}\int_0^t\!\!\!
dt'\phi^2_0(t',\vec{x})\int_0^{t'}\!\!\!dt''\phi^3_0(t'',\vec{x})\!+\!
O(\l^3)\; . \label{LIReqn2}\ee In the noninteracting (free)
theory, $\phi(t, \vec{x})=\phi_0(t, \vec{x})$. Hence, using
Eq.~(\ref{phizero}), the VEV of the scalar field-strength-squared
is obtained trivially in this ($\lambda=0$) limit \be \langle
\Omega \vert
\phi_0^2(x)\vert\Omega\rangle=\frac{H^2}{4\pi^2}\ln(a)\; .\ee This
stochastic result is the same as the results of Refs.~\cite {VF}
applying QFT.

\begin{figure}[htbp]
\centerline{\epsfig{file=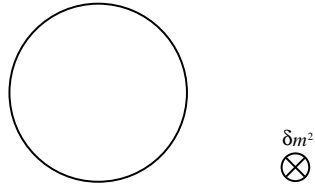,
                               width=1.6in,height=0.95in }}
\caption{Generic one-loop diagram with mass counterterm
\label{fig1} }
\end{figure}

\begin{figure}[htbp]
\centerline{\epsfig{file=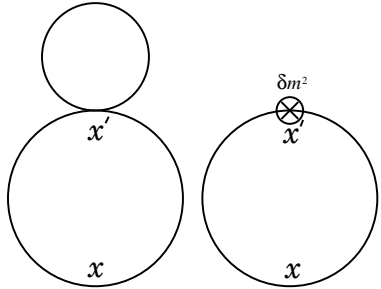,
                               width=1.95in,height=1.5in }}
\caption{Generic snowman diagram with mass counterterm.
\label{fig2} }
\end{figure}

Now we start calculating (i) the VEV $\lambda\langle \Omega \vert
\phi^2(x)\vert\Omega\rangle/2$ in the interacting $(\lambda\neq
0)$ theory. Using Eq.~(\ref{LIReqn2}) we find \be
\frac{\lambda}{2} \langle \Omega \vert
\phi^2(x)\vert\Omega\rangle\!\!&=&\!\!\frac{\lambda}{2}\!\left[\langle
\Omega \vert
\phi_0^2(x)\vert\Omega\rangle\!-\!\frac{\lambda}{9H}\!\!\int_0^t\!\!
dt' \langle\Omega\vert\phi_0(t, \vec{x})\phi_0^3(t',
\vec{x})\vert\Omega\rangle\right]\!\!+\!O(\l^3)\nonumber\\
\!\!&=&\!\!\frac{\l}{2}\!\left[\langle \Omega \vert
\phi_0^2(x)\vert\Omega\rangle\!-\!\frac{\lambda}{3H}\!\!\int_0^t\!\!
dt' \langle\Omega\vert\phi_0(t, \vec{x})\phi_0(t',
\vec{x})\vert\Omega\rangle\langle\Omega\vert\phi_0^2(t',
\vec{x})\vert\Omega\rangle\right]\!\!+\!O(\l^3)\;
.\hspace{0.8cm}\label{stomass}\ee Inserting Eq.~(\ref{phizero})
into Eq.~(\ref{stomass}) yields the stochastic result\be
\frac{\lambda}{2} \langle \Omega \vert
\phi^2(x)\vert\Omega\rangle\!\!&=&\!\!\frac{\l}{2}\!
\left[\frac{H^2}{4\pi^2}\ln(a)\!-\!\frac{\lambda}{3H}\!\!\int_0^t\!\!
dt'\frac{H^4}{16\,\pi^4}\ln^2(a')\!+\!O(\lambda^2)\right]\nonumber\\
\!\!&=&\!\!\frac{H^2}{2^3\pi^2}\lambda\ln(a)\left[1\!-\!\frac{1}{2^2
3^2\pi^2}\lambda\ln^2(a)\right]\!\!+\!O(\lambda^3)\; .
\label{MassSQ}
 \ee Next, we want to calculate the same VEV $\lambda
\langle\Omega|\phi^2|\Omega\rangle/2$ using QFT. Figure~\ref{fig1}
depicts the one-loop contribution. Hence, at one-loop order, the VEV is given in terms of
the coincident limit of the scalar propagator and the mass
counterterm $\delta m^2$:\be
\frac{\lambda}{2}\langle\Omega|\phi^2|\Omega\rangle=\frac{\lambda}{2}i\D(x;x)+\d
m^2 + O(\l^2) \; .\label{vev1l}\ee The scalar propagator in
$D$-dimensional locally de Sitter background is \cite{OW1,OW2}\be
i\D(x;x')=\frac{H^{D-2}}{(4\pi)^\frac{D}{2}}\Bigg\{-\sum_{n=0}^{\infty}
\frac{1}{n-\frac{D}{2}+1}\frac{\G(n+\frac{D}{2})}{\G(n+1)}
\left(\frac{y}{4}\right)^{n-\frac{D}{2}+1}
-\frac{\G(D-1)}{\G(\frac{D}{2})}\pi\cot{(\pi\frac{D}{2})}\nonumber\\
+\sum_{n=1}^{\infty}
\frac{1}{n}\frac{\G(n+D-1)}{\G(n+\frac{D}{2})}
\left(\frac{y}{4}\right)^n+\frac{\G(D-1)}{\G(\frac{D}{2})}\ln{(aa')}\Bigg\}\;
. \label{new} \ee Here the modified de Sitter length function
$y(x;x')$ is given in terms of the de Sitter conformal coordinate
interval $\Delta x^2$ (Eq.~(\ref{plusplus}))\beeq
y(x;x')=H^2aa'\Delta
x^2=H^2aa'\left[\|\vec{x}-\vec{x}'\|^2-(|\eta-\eta'|-i\delta)^2\right]\;
.\label{y}\eneq To facilitate dimensional regularization, we
express the dimension of spacetime in terms of its deviation from
four: $D=4-\epsilon$. Therefore, the coincident limit of the
scalar propagator\be i\D(x;x) = \lim_{x'\rightarrow x} i\D(x;x') =
\frac{H^{2-\e}}{(4\pi)^{2-\frac{\e}{2}}} \frac{\G(3 \!-\!
\e)}{\G(2 \!-\! \frac{\e}{2})} \Bigg\{2\ln(a) +
\pi\cot\Bigl(\frac{\pi\e}{2}\Bigr)\Bigg\} \; .\label{coin} \ee
Because of the finite, time-dependent term in Eq.~(\ref{coin}), we
cannot make the one-loop VEV~(\ref{vev1l}) vanish for all time.
Our renormalization condition is that it should be zero at $t=0$,
which implies \be \d m^2 = -\frac{\l
H^{2-\e}}{2^{5-\e}\pi^{2-{\frac{\e}2}}} \frac{\G\left(3 \!-\!
\e\right)}{\G\left(2\!-\! \frac{\e}2\right)} \pi
\cot\Bigl(\frac{\pi\e}{2}\Bigr) + O(\l^2) \; . \label{massct} \ee
Therefore, Eq.~(\ref{vev1l}) yields\beeq
\frac{\lambda}{2}\langle\Omega|\phi^2|\Omega\rangle_{\scriptscriptstyle\rm
1-loop} = \frac{\l H^2}{8\pi^2} \ln(a)\; . \label{sto1loop}\eneq
The two-loop diagram that contributes to the VEV is known as the
snowman diagram depicted on the left of Fig.
\ref{fig2}. The right-hand side
diagram depicts the one-loop mass counterterm which naturally
combines with it ($\d m^2$ denotes mass counterterm vertex). In
Schwinger-Keldish formalism (Sec.~\ref{sec:S-K}) the internal
vertices are summed over both $+$ and $-$ polarities. A simple
application of Feynman rules gives \be
\!\!\!\!\!\!\frac{\lambda}{2}\langle\Omega|\phi^2|\Omega\rangle_{\scriptscriptstyle\rm
2-loop} \!=\!\frac{\l}{2}\!\!\int\!\!d^Dx' a'^D\!\Biggl\{\!
[i\D_{++}(x;x')]^2\!-\![i\D_{+-}(x;x')]^2\!\Biggr\}\Bigg\{\!
\frac{(-i\l)}{2}i\D(x';x')\!-\!i\d m^2\!\Bigg\} .\label{snow}\ee
Both $++$ and $+-$ propagators are the same function (\ref{new})
of the appropriate version of the modified de Sitter length
function $y(x;x')$. By definition (\ref{plusplus}),
$y_{++}(x;x')\equiv H^2aa'\Delta x^2_{++}= y(x;x')$, given in
Eq.~(\ref{y}). On the other hand, $y_{+-}(x;x')\equiv H^2aa'\Delta
x^2_{+-}$, where the coordinate interval $\Delta x^2_{+-}$ is
given in Eq.~(\ref{plusminus}). The coincident propagator and the
mass counterterm are calculated in Eqs.~(\ref{coin}) and
(\ref{massct}), respectively. Because both diagrams in
Fig.~\ref{fig2} have the same lower loop, they posses the common
factor given in the first curly bracket of Eq.~(\ref{snow}). The
first term in the second curly bracket comes from the left-hand
side diagram, whereas the second term comes from the right-hand
side diagram.

The integral in Eq.~(\ref{snow}) is calculated explicitly in
Ref.~\cite{BOW}. The result can be read off directly from its
Eq.~(61). After renormalizing the overlapping divergence
$-\frac{\l^2
H^2}{2^7\pi^4}(\frac{2\pi}{H\mu})^\e\frac{\ln(a)}{\e}$ of the
snowman diagram by the two-loop mass counterterm, one obtains \be
&&\!\!\!\!\!\!\!\!\!\!\!\!\!\!\!\!\!\!\frac{\lambda}{2}\langle\Omega|\phi^2|
\Omega\rangle_{\scriptscriptstyle\rm 2-loop} \!=\!\frac{\l^2
H^2}{2^7\pi^4}\Bigg\{\!\!\!-\!\frac{4}{9}\ln^3(a)\!+\!\frac{13}{18}\ln^2(a)
\!+\!\!\left[\ln\!\left(\frac{H}{2\mu}\right)\!+\!\frac{8}{3}\!-\!\gamma
\!-\!\frac{2}{9}\pi^2\right]\!\ln(a)\!-\!\frac{238}{81}
\!+\!\frac{13}{54}\pi^2\nonumber\\
&&+\frac{4}{3}\zeta(3)
\!+\!\frac{a^{-3}}{81}\!-\!\sum_{n=1}^\infty\frac{n+5}{(n+1)^3}a^{-(n+1)}
\!+\!4\!\sum_{n=1}^\infty\frac{a^{-(n+2)}}{(n+2)^3}
\!+\!4\!\sum_{n=1}^\infty\frac{a^{-(n+3)}}{n(n+3)^3}\Bigg\}\!+\!
O(\l^3)\label{SMa}\; ,\ee where the Euler's constant $\gamma\simeq
0.577$. (The constants and exponentially decaying terms may be
subsumed into the definition of the vacuum state.) Combining one-
and two-loop results, Eqs.~(\ref{sto1loop}) and~(\ref{SMa}), we
find\beeq \frac{\lambda}{2}\langle\Omega|\phi^2|
\Omega\rangle=\frac{H^2}{2^3\pi^2}\lambda\ln(a)\left[1\!-\!\frac{1}{2^2
3^2\pi^2}\lambda\ln^2(a)\right]\!\!+\!O(\lambda^3)\; ,
\label{qftvev}\eneq in leading logarithm order. This QFT result is
exactly the same as stochastic result Eq.~(\ref{MassSQ}).

Until the breakdown of the perturbation theory ---that occurs
around $\lambda\ln^2(a)\sim 1$, as was pointed out earlier---
expectation value~(\ref{MassSQ}) remains positive. This means
that, the curvature associated with the scalar being away from the
minimum of the potential assumes a growing positive expectation
value which {\it acts like} a positive ``mass-squared'' during the
process. This agrees with the decreasing mode function obtained in
Eq.~(\ref{Modefunc}) by solving the one- and two-loop corrected
effective field equation, in the context of QFT.

Next, in calculation (ii), we recompute the very same mode
solution $\Phi(x;\vec{k})$ obtained in Eq.~(\ref{Modefunc}) using
QFT. This time, however, we calculate
$\Phi(x;\vec{k})=\langle\Omega\vert[\phi(x),
\alpha^{\dagger}(\vec{k})]\vert \Omega\rangle$
(Eq.~(\ref{calphi})) by applying the stochastic technique to show
that the two realizations yield exactly the same result, in
leading logarithm order. The commutator \be \Bigl[\phi(x),
\alpha^{\dagger}(\vec{k})\Bigr]\!\!\!&=&\!\!\!u(0,k)e^{i\vec{k}\cdot\vec{x}}\Biggl\{1\!-\!
\frac{\l}{6H}\int_0^t\!\!\!dt' \phi^2_0(t',
\vec{x})\!+\!\frac{\l^2}{2^2\,3^2H^2}
\int_0^t dt'\phi^2_0(t',\vec{x}) \int_0^{t'}\!\!\!dt''\phi_0^2(t'',\vec{x}) \nonumber \\
&&\hskip 3.2cm+\frac{\l^2}{2\,3^3 H^2}
\int_0^t\!\!\!dt'\phi_0(t',\vec{x})
\int_0^{t'}\!\!\!dt''\phi^3_0(t'',\vec{x})\Biggr\}\!+\!O(\l^3)  \;
,\label{phiLIR1}\ee which implies\be
\Phi(x;\vec{k})\!\!&=&\!\!\Bigl\langle \Omega \Bigl\vert
\Bigl[\phi(x), \alpha^{\dagger}(\vec{k})\Bigr] \Bigr\vert \Omega
\Bigr\rangle = u(0,k)
e^{i\vec{k}\cdot\vec{x}}\Biggl\{1\!-\!\frac{\l}{6H}\!\int_0^t\!\!
dt'\langle\Omega\vert\phi^2_0(t',
\vec{x})\vert\Omega\rangle\nonumber\\
&&+\frac{\l^2}{2^2\; 3^2
H^2}\Bigg[\int_0^t\!\!dt'\!\!\int_0^{t'}\!\!\!dt''\langle\Omega\vert\phi^2_0(t',
\vec{x})\vert\Omega\rangle\langle\Omega\vert\phi^2_0(t'',
\vec{x})\vert\Omega\rangle\nonumber\\&&+2\int_0^t\!\!dt'\!\!\int_0^{t'}
\!\!\!dt''\Bigl(\langle\Omega\vert\phi_0(t', \vec{x})\phi_0(t'',
\vec{x})\vert\Omega\rangle\Bigr)^2\Bigg]\nonumber\\&&+
\frac{\l^2}{2\, 3^2 H^2} \int_0^t\!\!dt'\!\!\int_0^{t'}\!\!
dt''\langle\Omega\vert\phi_0(t', \vec{x})\phi_0(t'',
\vec{x})\vert\Omega\rangle\langle\Omega\vert\phi^2_0(t'',
\vec{x})\vert\Omega\rangle
 \Biggr\}+ O(\l^3) \; .\label{VEV}\ee Using Eq.~(\ref{phizero}) in Eq.~(\ref{VEV}),
 yields
\be\Phi(x;\vec{k})\!\!&=&\!\!
u(0,k)e^{i\vec{k}\cdot\vec{x}}\Biggl\{\!1\!-\!\frac{\l}{6H}\!\int_0^t\!\!
dt' \frac{H^2}{4\pi^2}\ln(a')\hspace{4.8cm}\nonumber\\ &&\!\!+
\frac{\l^2}{2^2 3^2 H^2}
\!\int_0^t\!\!\!\!dt'\frac{H^2}{4\pi^2}\ln(a')
\!\int_0^{t'}\!\!\!\!
dt''\frac{H^2}{4\pi^2}\ln(a'')\!+\!\frac{\l^2}{ 3^2 H^2}
\int_0^t\!\!\!\!dt'\!\int_0^{t'}\!\!\!\!dt''
\frac{H^4}{16\,\pi^4}\ln^2(a'')\Biggr\}\!+\!O(\l^3)\nonumber\\
\!\!&=&\!\!u(0,k)e^{i\vec{k}\cdot\vec{x}}\Bigg\{\!1\!-\!\frac{1}{2^4
3\pi^2}\lambda\ln^2(a)\!+\!\frac{11}{2^9 3^3\pi^4}\l^2\ln^4(a)
\Bigg\}+ O(\l^3)\; .\hspace{0.5cm}\label{Modefunc2}\ee This result
is in perfect agreement with Eq.~(\ref{Modefunc}), which is
obtained by lengthy and highly nontrivial calculation using
quantum field theory.

\section{Conclusions}
\label{sec:Conc}

Massless, minimally coupled $\frac{\lambda}{4!}\varphi^4$ on a
locally de Sitter background can induce enhanced quantum effects
causing super-accelerated phase of cosmic expansion, a possibility
not excluded by present observations. In this paper, we have
studied the stability of this system for $\lambda\ll 1$. In
Sec.~\ref{sec:EffectiveMode}, we have obtained the
quantum-corrected effective field equations at linearized order,
using the fully renormalized Schwinger-Keldish self-mass-squared
terms at one- and two-loop orders. In Sec.~\ref{sec:modefunc}, we
have solved the effective field equations in the late time limit,
i.e., for $\ln(a)\gg 1$, and obtained the scalar mode function in
leading powers of infrared logarithms at each order of
perturbation. In Sec.~\ref{sec:Stoch}, we have used Starobinsky's
stochastic inflation technique to compute the mode function in the
leading logarithm approximation and compared it with the quantum
field theory result of Sec.~\ref{sec:modefunc}.

There are three main conclusions that we draw: (i) perturbation
theory breaks down for $\ln(a(t))\sim 1/\sqrt{\l}$. This, however,
does not invalidate the reliability of our late time solutions
since one can have long period of time during which $1\ll\ln(a)\ll
1/\sqrt{\lambda}$ for $\lambda\ll1$. (ii) The quantum-corrected
mode function decreases in time ---consistent with the field
developing a positive (nontachyonic) mass-squared--- starting from
its initial classical (Bunch-Davies) value. This means that linear
perturbations do not grow in this model. Thus, the model is
stable. (iii) The results of quantum field theory and
Starobinsky's stochastic technique are in perfect agreement.

The effect can be understood physically as follows. Heisenberg's
energy-time uncertainty principle implies that {\it virtual}
particle-antiparticle pairs continually emerge from the vacuum and
then disappear back into it. However, massless particles which are
also conformally noninvariant have a certain amplitude for
appearing with wavelength greater than the inverse of the Hubble
parameter. In an {\it inflating} universe, these virtual pairs are
pulled apart by the Hubble flow before they find time to
annihilate each other. Hence, they become real, recalling the
analogy with the Hawking radiation. Therefore, one gains particles
out of nothing which also means that the scalar field strength
{\it grows}. In fact $\langle\Omega|\varphi^2|\Omega\rangle=
\frac{H^2}{4\pi^2}Ht+O(\lambda)$ in our model. Thus, inflationary
particle production causes the scalar to undergo a random walk
such that its average distance from the minimum of the quartic
potential increases. This makes the vacuum energy larger; hence
$\dot\rho>0$. Now recall the covariant stress-energy conservation
law $\dot\rho=-3H(\rho+p)$. Because inflationary particle
production causes $\dot\rho>0$, we must have $\rho+p<0$ to satisfy
this law. Hence, the Weak Energy Condition is violated and the
equation of state parameter $w<-1$. Will this effect be
terminated? If the growing of the field, which generates the
effect, {\it stops}, then the effect terminates. There are two
causes, in the interacting theory, which would yield the growing
of the field come to a halt eventually. The first cause is the
{\it classical restoring force} $-\frac{\lambda}{6}\varphi^3$.
This force pushes the field back down to the configuration where
the potential is minimum, i.e., to $\varphi=0$, as the scalar
rises up its potential. The second cause is the curvature
associated with field being away from the minimum of the
potential, i.e., $\frac{\lambda}{2}\varphi^2$ which acts like a
``mass-squared.'' Because
$\frac{\lambda}{2}\langle\Omega|\varphi^2|\Omega\rangle$ is a
growing positive real number, the scalar develops a growing
positive mass. That cuts off particle production, since
inflationary particle production requires effective masslessness
(with respect to the Hubble parameter). Because of these two
causes, the field can not continue to roll up its position. It
must eventually come to a halt. In fact, Starobinsky and Yokoyama
showed \cite{SY} that $\langle\Omega|\varphi^2|\Omega\rangle$ does
asymptote to a positive {\it constant}.

\begin{appendix}

\section{Integrating the two-loop terms}
\label{App:secondint}

In this appendix we calculate the integral \be
&&\!\!\!\!\!\!\!\!\!\!\!\!\!\!\!\!\!\!-\int_{\eta_i}^0\!
d\eta'\!\int\! d^3x'\! \mathcal{M}_2^2(x;x')u(\eta', k)e^{-i
\vec{k} \cdot(\vec{x}- \vec{x}\,')}\; ,\label{M2A}\ee which is a
part of Eq.~(\ref{IntegroPhi2}). $\mathcal{M}_2^2(x;x')$ is
defined in Eq.~(\ref{2el}). In the late time regime of physical
interest, we replace $u(\eta', k)$ (Eq.~(\ref{u})) with its
constant limit $u(0,k)$. Moreover, as can be seen below, when $++$
and $+-$ terms are added, factors of the Heaviside function
$\Theta(\D\eta-\D r)$ arise. They require
$\|\vec{x}-\vec{x}'\|\leq\eta-\eta'$, ensuring casuality in the
Schwinger-Keldish formalism. In the late time limit this means
that the spatial plane wave factor
$e^{i\vec{k}\cdot(\vec{x}-\vec{x}')}$ can also be dropped in
Eq.~(\ref{M2A}), since $\eta\!-\!\eta'\rightarrow 0$ in that
regime. We, therefore, break the integral (\ref{M2A}) into a sum
of six terms of the general form $\sum_{n=1}^6 I_n(\eta)$, with
\be I_n(\eta)\equiv-u(0, k)\!\int_{\eta_i}^0\! d\eta'\!\int\!
d^3x'\!\mathcal{M}_{2, n}^2(x;x')\label{I_n}\; .\ee Here, we
define the integrands \be
\!\!\!&&\!\!\!\!\!\!\!\!\!\mathcal{M}^2_{2, 1}(x;x')= \frac{i
}{2^{12}\,3\pi^6} a a' \dd^4 \Biggl[\frac{\ln\left(\mu^2 \D
x^2_{\scriptscriptstyle ++}\right)}{\D x^2_{\scriptscriptstyle
++}} - \frac{\ln\left(\mu^2 \D x^2_{\scriptscriptstyle
+-}\right)}{\D x^2_{\scriptscriptstyle +-}}\Biggr]\label{m21}\; ,\\
&& \!\!\!\!\!\!\!\!\!\mathcal{M}^2_{2, 2}(x;x')=-\frac{i H^2}{2^9
\pi^6}\! (a a')^2 \dd^2 \Biggl[ \ln\Bigl(\frac{H e^{\frac34}}{2
\mu}\!\Bigr) \Bigg(\frac{\ln\left(\mu^2 \D x^2_{
\scriptscriptstyle ++}\right)}{\D x^2_{\scriptscriptstyle ++}}-
\frac{\ln\left(\mu^2 \D x^2_{ \scriptscriptstyle +-}\right)}{\D
x^2_{\scriptscriptstyle +-}}\Bigg)\Biggr]\label{m22}\; ,\\
&& \!\!\!\!\!\!\!\!\!\mathcal{M}^2_{2, 3}(x;x')=-\frac{i
H^2}{2^{11} \pi^6}\! (a a')^2 \dd^2 \Biggl[\frac{\ln^2\left(\mu^2
\D x^2_{\scriptscriptstyle ++}\right)}{\D x^2_{ \scriptscriptstyle
++}}-\frac{\ln^2\left(\mu^2 \D x^2_{\scriptscriptstyle
+-}\right)}{\D x^2_{ \scriptscriptstyle +-}}\Biggr]\label{m23}\; ,\\
&&\!\!\!\!\!\!\!\!\!\mathcal{M}^2_{2, 4}(x;x')= -\frac{i H^4}{2^9
\pi^6}\! (a a')^3 \Bigg[\frac{ \ln^2\left(\frac{\sqrt{e}}4 H^2 \D
x^2_{\scriptscriptstyle ++}\right)}{\D x^2_{\scriptscriptstyle
++}}- \frac{ \ln^2\left(\frac{\sqrt{e}}4 H^2 \D
x^2_{\scriptscriptstyle +-}\right)}{\D x^2_{\scriptscriptstyle
+-}}\Bigg]\label{m24}\; ,\\
&&\!\!\!\!\!\!\!\!\!\mathcal{M}^2_{2, 5}(x;x')= \frac{i
H^6}{2^{10}\,3\pi^6}(a a')^4\!\Bigg[ \ln^3\Bigl(\frac{\sqrt{e}}4
H^2 \D x^2_{ \scriptscriptstyle ++}
\Bigr)-\ln^3\Bigl(\frac{\sqrt{e}}4
H^2 \D x^2_{ \scriptscriptstyle +-} \Bigr)\Bigg]\label{m25}\; ,\\
&&\!\!\!\!\!\!\!\!\!\mathcal{M}^2_{2, 6}(x;x')=-\frac{1}{2^9 3\,
\pi^4} \, a^2 \Biggl\{\ln(a) \dd^2 \!-\! \Bigl(2 \ln(a) \!+\!
1\Bigr) H a
\partial_0 \Biggr\} \delta^4(x \!-\! x')\nonumber\\&&\!\!\!\!\!\!
- \frac{H^2}{2^7\,3^2\pi^4}a^4 \Biggl\{4 \ln^3(a) \!+\!
\frac{23}{2} \ln^2(a)\!-\! \Bigl[39 \!+\! 27 \ln\Bigl(\frac{H}{2
\mu} \Bigr) \!-\! 2 \pi^2 \Bigr] \ln(a) \Biggl\} \delta^4(x \!-\!
x')\nonumber\\&&\!\!\!\!\!\!+ \frac{H^2}{2^7 \pi^4}a^4 \Biggl\{
\frac{a^{-3}}{81} \!-\! \sum_{n=1}^\infty
\frac{n+5}{(n+1)^3}a^{-(n+1)} \!+\! 4 \sum_{n=1}^\infty
\frac{a^{-(n+2)}}{(n+2)^3}\!+\! 4
\sum_{n=1}^\infty\frac{a^{-(n+3)}}{n(n+3)^3} \Biggr\} \delta^4(x
\!-\! x') \label{m26}\; .\ee We evaluate $I_1(\eta)$ and
$I_3(\eta)$ explicitly to illustrate the relevant calculation
techniques and give only the final answers for the remaining four
that can be obtained similarly. The first integral is\be
I_1(\eta)\equiv-u(0, k)\int_{\eta_i}^\eta d\eta'\int
d^3x'\mathcal{M}_{2, 1}^2(x;x')\label{apI1}\; .\ee It is useful to
break up the logarithms in $\mathcal{M}_{2, 1}^2$ as
\beeq\ln(\mu^2 \D x^2)=\ln\Bigl(\frac{H^2\D
x^2}{4}\Bigr)+2\ln\Bigl(\frac{2\mu}{H}\Bigr)\label{lnmux^2}\;
.\eneq We then partially integrate the inverse powers of $\D x^2$,
using the identities \be \frac{ \ln\left(H^2 \D x^2\right)}{\D
x^2}&=&\frac{\dd^2}{8}\ln^2\left(H^2 \D x^2\right)-\frac{1}{\D
x^2}\; ,\label{ln}\\
\frac{1}{\D x^2}&=&\frac{\dd^2}{4}\ln\left(H^2 \D
x^2\right)\label{1/x}\; .\ee Because the integral is over
$x'^\mu$, the derivatives with respect to $x^\mu$ can be taken
outside the integral. The remaining integrand possesses only
logarithmic singularities. There is no distinction between the
$++$ and $+-$ terms at this stage.  We define the temporal and
spatial intervals $\D\eta\equiv \eta-\eta'$ and $\D
r\equiv\|\vec{x}-\vec{x}'\|$, respectively. The $++$ and $+-$
terms cancel for $\D\eta<0$, so we can restrict the integration to
$\D\eta>0$. Then, the logarithms can be expanded as
\be\ln\Bigl(\frac{H^2}{4}\D x^2_{+\pm}\Bigr)&=&
\ln\Bigl(\frac{H^2}{4}(\D\eta^2\!-\!\D r^2)\Bigr)\pm
i\pi\Theta(\D\eta^2\!-\!\D r^2)\;. \label{defnlog}\ee  Making the
change of variables $\vec{r}=\vec{x}-\vec{x}'$ and performing the
angular integrals yield \be I_1(\eta)=\!-\frac{1}{2^{11}\,3\pi^4}
u(0,k) a\, \dd_0^6 \int_{\eta_i}^{\eta}\! d\eta'
a'\!\int_0^{\D\eta}\! dr\, r^2
\Bigl[\ln\Bigl(\frac{H^2}{4}(\D\eta^2\!-\!\D r^2)
\Bigr)\!+\!2\ln\Bigl(\frac{2\mu}{H}\Bigr)\!-\!1\Bigr]\;
,\nonumber\ee where the initial time $\eta_i\equiv -H^{-1}$. Next,
we make the change of variables $r\equiv\D\eta\, z$ and perform
the integration over $z$, using \be \int_0^1 dz z^2
\ln\Bigl(\frac{1 \!-\! z^2}4\Bigr) & = & - \frac89 \;
.\label{x^2ln}\ee The result is \be I_1(\eta)=\!-\frac{
1}{2^{10}\,3^2\pi^4} u(0,k) a\, \dd_0^3 \int_{\eta_i}^{\eta}\!
d\eta' a'\! \dd_0^3\Bigl[\D\eta^3\Bigl(\ln(H\D\eta)
\!+\!\ln\Bigl(\frac{2\mu}{H}\Bigr)\!-\!\frac{11}{6}\Bigr)\Bigr]\;
.\ee Note that, owing to the factor $\D\eta^3$, three of the
external derivatives were brought inside the integral. This cubic
derivative gives $6\ln(2\mu\D\eta)$, when it acts upon the terms
inside the square bracket. At this stage, one makes the change of
variables $\eta'=-(Ha')^{-1}$ and looks up the relevant integral,
Eq.~(\ref{1/aln}), from Appendix~\ref{App:Bintegrals}. Acting on
the remaining derivatives using $\dd_0=Ha^2\frac{\dd}{\dd a}$, one
obtains \be
I_1(\eta)\!\!&=&\!\!\frac{H^2}{2^8\,3\pi^4}u(0,k)a^4\Bigg\{\ln(a)
-\ln\Bigl(\frac{2\mu}{H}\Bigr)+\frac{3}{2}
+\sum_{n=1}^\infty\frac{(n-1)(n-2)}{2n}a^{-n}\Bigg\}\; .\ee
Evaluation of $I_2(\eta)$, defined by Eqs.~(\ref{I_n}) and
(\ref{m22}), is similar to that of $I_1(\eta)$. Using
Eqs.~(\ref{lnmux^2})-(\ref{x^2ln}) and (\ref{a^0ln}) yields\be
I_2(\eta)\!\!&=&\!\!\frac{H^2}{2^6
\pi^4}\ln\Bigl(\frac{He^\frac{3}{4}}{2\mu}\Bigr)u(0,k)a^4
\Bigg\{\ln(a) -\ln\Bigl(\frac{2\mu}{H}\Bigr)+1-\ln\Bigl(1 \!-\!
\frac1{a}\Bigr) -a^{-1}\Biggr\}\;.\ee Next, we explicitly evaluate
\be I_3(\eta)\equiv-u(0, k)\int_{\eta_i}^\eta d\eta'\int
d^3x'\mathcal{M}_{2, 3}^2(x;x')\; ,\label{apI3}\ee where
$\mathcal{M}_{2, 3}^2$ is given in Eq.~(\ref{m23}). We break up
the logarithm squared in $\mathcal{M}_{2, 3}^2$ as\beeq\ln^2(\mu^2
\D x^2)=\ln^2\Bigl(\frac{H^2\D
x^2}{4}\Bigr)+4\ln\Bigl(\frac{2\mu}{H}\Bigr)\ln\Bigl(\frac{H^2\D
x^2}{4}\Bigr)+4\ln^2\Bigl(\frac{2\mu}{H}\Bigr)\label{break}\;
.\eneq Then, we use the identity \be \frac{\ln^2\left(H^2 \D
x^2\right)}{\D x^2}&=&\frac{\dd^2}{12}\ln^3\left(H^2 \D
x^2\right)-\frac{\dd^2}{4}\ln^2\left(H^2 \D x^2\right)+\frac{2}{\D
x^2}\label{ln^2} \; \ee and Eqs.~(\ref{1/x}), (\ref{defnlog}). To
evaluate the radial integral, we make the change of variables
$r\equiv \D\eta z$ and use Eq.~(\ref{x^2ln}) and the integral
\be\int_0^1 dz z^2 \ln^2\Bigl(\frac{1 \!-\! z^2}4\Bigr) & = &
\frac{104}{27} - \frac{\pi^2}9\label{x^2ln^2}\; .\ee We find \be
I_3(\eta)=\!-\frac{H^2}{2^8\,3\pi^4} u(0,k) a^2\, \dd_0
\int_{\eta_i}^{\eta}\! d\eta' a'^2\! \dd_0^3\Bigl[
\D\eta^3\Bigl(\ln^2(H\D\eta) \!+\!A\ln(H\D\eta) +B\Bigr)\Bigr]\;
,\ee where $A\equiv2\ln\left({2\mu}/{H}\right)\!-\!{11}/{3}$ and
$B\equiv\ln^2\left({2\mu}/{H}\right)
-({11}/{3})\ln\left({2\mu}/{H}\right)+{85}/{18}-{\pi^2}/{6}$. The
cubic derivative in the integrand yields
$6\ln^2(2\mu\D\eta)-\pi^2$, when it acts upon the terms inside the
square bracket. Making the change of variables $\eta'=-(Ha')^{-1}$
and using Eqs.~(\ref{a^0ln})-(\ref{a^0ln^2}) and
$\dd_0=Ha^2\frac{\dd}{\dd a}$, one obtains \be
I_3(\eta)\!\!&=&\!\!\frac{H^2}{2^6 \pi^4}u(0,k)a^4
\Bigg\{\!\!-\!\frac{1}{2}\ln^2(a)
\!+\!\Bigl[\ln\Bigl(\frac{2\mu}{H}\Bigr)\!-\!1\Bigr]\ln(a)-
\Bigl[\frac{1}{2}\ln\Bigl(\frac{2\mu}{H}\Bigr)
\!-\!1\!+\!a^{-1}\Bigr]\ln\Bigl(\frac{2\mu}{H}\Bigr)\!-\!\frac{\pi^2}{12}
\nonumber\\&&\hspace{2.5cm}-\Bigl[\ln\Bigl(\frac{2\mu}{H}\Bigr)
\!+\!a^{-1}\Bigr]\ln\Bigl(1\!-\!
\frac1{a}\Bigr)\!-\!\frac{1}{2}\ln^2\Bigl(1\!-\! \frac1{a}\Bigr)
\!-\!\sum_{n=1}^\infty\frac{(n-1)}{n^2}a^{-n}\Bigg\}\; .\ee
Evaluation of $I_4(\eta)$, defined by Eqs.~(\ref{I_n}) and
(\ref{m24}), is similar to that of $I_3(\eta)$. Using
Eqs.~(\ref{1/x})-(\ref{x^2ln}), (\ref{break})-(\ref{x^2ln^2}),
(\ref{a^0ln}), (\ref{a^1ln}), (\ref{a^0ln^2}) and (\ref{a^1ln^2})
yields \be
&&\!\!\!\!\!\!\!\!\!\!\!\!\!\!I_4(\eta)\!=\!\frac{H^2}{2^6
\pi^4}u(0,k)a^4
\Bigg\{\ln^2(a)-\frac{1}{2}\ln(a)\!+\!\frac{1}{16}\!+\!\frac{\pi^2}{6}
\!+\!\Bigl[\frac{1}{2}\!+\!a^{-1}\!-\!\frac{3}{2}a^{-2}\Bigr]\ln\Bigl(1\!-\!
\frac1{a}\Bigr)\nonumber\\
&&\!\!\!\!\!\!\!\!\!\!\!\!+\Bigl[1\!-\!2a^{-1}\!+\!a^{-2}\Bigr]\ln^2\Bigl(1\!-\!
\frac1{a}\Bigr)\!-\!\Bigl[\frac{13}{8}
-\frac{\pi^2}{3}\Bigr]a^{-1}+\Bigl[\frac{25}{16}
-\frac{\pi^2}{6}\Bigr]a^{-2}-2\sum_{n=1}^\infty\frac{a^{-n}}{n^2}\Bigg\}\;
.\ee To evaluate $I_5(\eta)$, we expand $\ln^3\Bigl(\sqrt{e}H^2 \D
x^2/4 \Bigr)=\left[\ln\Bigl(H^2 \D x^2/4 \Bigr)+(1/2)\right]^3$
and use Eqs.~(\ref{defnlog}), (\ref{x^2ln}), (\ref{x^2ln^2}) and
(\ref{1/aln^2})-(\ref{a^2ln^2}). The result is\be
&&\!\!\!\!\!\!\!\!\!\!\!\!\!\!\!\!\!\!\!I_5(\eta)\!=\!\frac{H^2}{2^5\,
3^2\pi^4}u(0,k)a^4
\Biggl\{\ln^3(a)-\frac{9}{4}\ln^2(a)+\Bigl[\frac{15}{16}+\frac{\pi^2}{2}\Bigr]\ln(a)
\!-\!\frac{2035}{288}\!+\!\frac{\pi^2}{6}\nonumber\\
&&\!\!\!\!\!\!\!\!\!\!\!\!\!\!\!\!\!\!\!+\Bigl[\frac{71}{12}\!-\!\frac{25}{2}a^{-1}
\!+\!\frac{35}{4}a^{-2}\!
-\!\frac{13}{6}a^{-3}\Bigr]\ln\Bigl(1\!-\!\frac1{a}\Bigr)-\Bigl[\frac{11}{2}\!-\!9a^{-1}
\!+\!\frac{9}{2}a^{-2}\!-\!a^{-3}\Bigr]\ln^2\Bigl(1\!-\!
\frac1{a}\Bigr)\nonumber\\
&&\!\!\!\!\!\!\!\!\!\!\!\!\!\!\!\!\!\!\!+\Bigl[\frac{671}{48}\!-\!\frac{3\pi^2}{2}\Bigr]a^{-1}\!
-\!\Bigl[\frac{883}{96}
\!-\!\frac{3\pi^2}{4}\Bigr]a^{-2}\!+\!\Bigl[\frac{329}{144}
\!-\!\frac{\pi^2}{6}\Bigr]a^{-3}\!-\!6\sum_{n=1}^\infty\frac{a^{-n}}{n^2}
\Bigl[\psi(n)\!+\!\gamma\!-\!\frac{3}{4}\!-\!\frac{1}{n}\,\Bigr]\Biggr\}\;
, \ee where the Digamma function
$\psi(n)\equiv-\gamma+\sum_{k=1}^{n-1}k^{-1}$ and Euler's gamma
$\gamma\simeq 0.577$. It is straightforward to show that the
remaining integral \be I_6(\eta)\!\!&=&\!\!\frac{H^2}{2^5 3^2
\pi^4}u(0,k)a^4 \Biggl\{\ln^3(a)\!+\!\frac{23}{8}\ln^2(a)
\!+\!\Bigl[\frac{27}{4}\ln\Bigl(\frac{2\mu}{H}\Bigr)
\!-\!\frac{39}{4}\!+\!\frac{\pi^2}{2}
\Bigr]\ln(a)\!-\!\frac{1}{36}a^{-3}\nonumber\\
&&\hspace{2.3cm}\!+\!\frac{9}{4}\sum_{n=1}^{\infty}\frac{n+5}{(n+1)^3}a^{-(n+1)}
\!-\!9\sum_{n=1}^{\infty}\frac{a^{-(n+2)}}{(n+2)^3}
\!-\!9\sum_{n=1}^{\infty}\frac{a^{-(n+3)}}{n(n+3)^3}\Bigg\}\; .\ee
Summing the six terms gives the total two-loop contribution in
Eq.~(\ref{IntegroPhi2})\be
&&\!\!\!\!\!\!\!\!\!\!\!\!\!\!\!\!\!\!-\int_{\eta_i}^\eta
d\eta'\int d^3x' \mathcal{M}_2^2(x;x')u(\eta', k)e^{-i \vec{k}
\cdot(\vec{x}-
\vec{x}\,')}\longrightarrow \sum_{n=1}^6 I_n(\eta)\nonumber\\
&&\!\!\!\!\!\!\!\!\!\!\!\!\!\!\!\!\!\!=\frac{H^2}{2^4\,3^2\pi^4}u(0,k)a^4\Bigg\{\ln^3(a)
\!+\!\frac{23}{16}\ln^2(a)\!+\!\left[\frac{27}{8}\ln\Bigl(\frac{2\mu}{H}\Bigr)
\!-\!\frac{189}{32}\!+\!\frac{\pi^2}{2}\right]\ln(a)\nonumber\\
&&\!\!\!\!\!\!\!\!\!\!\!\!\!\!+\!\left[\frac{9}{8}\ln\Bigl(\frac{2\mu}{H}\Bigr)
\!-\!\frac{15}{8}\right]\ln\Bigl(\frac{2\mu}{H}\Bigr)\!-\!\frac{205}{144}
\!+\!\frac{13}{48}\pi^2\!+\!\left[\frac{115}{48}\!-\!\frac{25}{4}a^{-1}\!+\!a^{-2}
\!-\!\frac{13}{12}a^{-3}\right]\ln\Bigl(1\!-\!\frac{1}{a}\Bigr)\nonumber\\
&&\!\!\!\!\!\!\!\!\!\!\!\!\!\!-\!\left[\frac{13}{8}\!-\!\frac{a^{-3}}{2}\right]
\ln^2\Bigl(1\!-\!\frac{1}{a}\Bigr)\!+\!\frac{79}{48}
a^{-1}\!-\!\frac{13}{12}a^{-2}
\!+\!\left[\frac{325}{288}\!-\!\frac{\pi^2}{12}\right]\!a^{-3}
\!+\!\frac{9}{8}\sum_{n=1}^\infty\!\frac{n\!+\!5}{(n\!+\!1)^3}a^{-(n+1)}\nonumber\\
&&\!\!\!\!\!\!\!\!\!\!\!\!\!\!
-\frac{9}{2}\sum_{n=1}^\infty\frac{a^{-(n+2)}}{(n\!+\!2)^3}
\!-\!\frac{9}{2}\sum_{n=1}^\infty\frac{a^{-(n+3)}}{n(n\!+\!3)^3}
\!+\!\frac{3}{8}\sum_{n=1}^\infty\left[\frac{n}{4}\!-\!\frac{3}{4}\!-\!\frac{11}{2n}
\!-\!\frac{8}{n^2}\left(\psi(n)\!+\!\gamma\!-\!\frac{1}{n}\right)\right]\!a^{-n}
\!\Bigg\}\; .\ee

\section{Useful Integral Identities}
\label{App:Bintegrals}

In Appendix \ref{App:secondint}, to calculate the temporal
integrations over $\eta'$, we make the change of variables
$\eta'=-(Ha')^{-1}$, and use the following integral identities:
\begin{eqnarray}
\int_1^a \frac{da'}{a'} \ln\Bigl(\frac1{a'} - \frac1{a}\Bigr) &=&
-\frac12 \ln^2(a) \!-\! \sum_{n=1}^{\infty} \frac{(1 \!-\! a^{-n})}{n^2} , \label{1/aln}\\
\int_1^a da' \ln\Bigl(\frac1{a'} - \frac1{a}\Bigr) & = & - a\ln(a)
+ (a - 1)\ln\Bigl(1 \!-\! \frac1{a}\Bigr)\; , \label{a^0ln}\\
\int_1^a da' a' \ln\Bigl(\frac1{a'} - \frac1{a}\Bigr)&=& -
\frac{a^2}{2} \ln(a) \!-\! \frac12 (a^2 \!-\! a) \!+\! \frac12
(a^2 \!-\! 1) \ln\Bigl(
1 \!-\! \frac1{a}\Bigr) \;,\label{a^1ln}\\
\int_1^a da' a'^2 \ln\Bigl(\frac1{a'} - \frac1{a}\Bigr)&=& -
\frac{a^3}{3} \ln(a) \!-\! \frac{a^3}{2} \!+\! \frac{a^2}{3} \!+\!
\frac{a}{6}+\frac{1}{3}(a^3-1) \ln\Bigl(
1 \!-\! \frac1{a}\Bigr)\; ,\label{a^2ln} \\
\int_1^a \frac{da'}{a'} \ln^2\Bigl(\frac1{a'} - \frac1{a}\Bigr)&=&
\frac13 \ln^3(a) \!+\! \frac{\pi^2}3 \ln(a) \!+\! 2
\sum_{n=1}^{\infty} \frac{(1 \!-\! a^{-n})}{n^2} \Bigl[ \psi(n)
\!+\! \gamma \!-\! \frac1{n}\Bigr] \; ,\label{1/aln^2}\ee where
$\psi(n)$ is the Digamma function and $\gamma$ is the Euler's
constant, as defined in Appendix~\ref{App:secondint}, \be\int_1^a
da' \ln^2\Bigl(\frac1{a'} - \frac1{a}\Bigr) & = & a \ln^2(a) + 2 a
\sum_{n=1}^{\infty} \frac{(1-a^{-n})}{n^2} + (a-1)
\ln^2\Bigl(1 \!-\! \frac1{a}\Bigr)\; ,\label{a^0ln^2}\\
\int_1^a da' a' \ln^2\Bigl(\frac1{a'} - \frac1{a}\Bigr) &=&
\frac{a^2}{2} \ln^2(a) \!+\! a^2 \ln(a) \!+\! a^2
\sum_{n=1}^{\infty} \frac{(1 \!-\! a^{-n})}{n^2}- a (a \!-\! 1)
\ln\Bigl(1 \!-\! \frac1{a}\Bigr) \nonumber\\ &&\hspace{5.5cm} +
\frac12 (a^2
\!-\! 1) \ln^2\Bigl(1 \!-\! \frac1{a}\Bigr) \; ,\label{a^1ln^2}\\
\int_1^a da' a'^2 \ln^2\Bigl(\frac1{a'} - \frac1{a}\Bigr) &=&
\frac{a^3}{3} \ln^2(a) \!+\! a^3 \ln(a) \!+\!
\frac{a^2}{3}(a-1)+\frac{2}{3} a^3 \sum_{n=1}^{\infty} \frac{(1
\!-\! a^{-n})}{n^2}\nonumber\\ && \hspace{0.5 cm} - \frac{a}{3} (a
\!-\! 1)(3a+1) \ln\Bigl(1 \!-\! \frac1{a}\Bigr) + \frac13 (a^3
\!-\! 1) \ln^2\Bigl(1 \!-\! \frac1{a}\Bigr) \; .\label{a^2ln^2}\ee
\end{appendix}
\begin{acknowledgements}
The authors thank Richard Woodard for his guidance and stimulating
discussions. EOK is grateful to Nikolaos Tsamis for his
hospitality during a recent visit to the University of Crete. EOK
was supported by NSF grant PHY-0244714 and by the Institute for
Fundamental Theory at the University of Florida. VKO was supported
by European Commission Marie Curie Fellowship FP-6-012679 and by
the Institute of Plasma Physics at the University of Crete.
\end{acknowledgements}

\end{document}